# Efficiency limits of solar energy harvesting via internal photoemission in carbon materials

**Svetlana V. Boriskina**[*], **Jiawei Zhou, Zhiwei Ding, and Gang Chen**[*]

Department of Mechanical Engineering, Massachusetts Institute of Technology, Cambridge, MA 02139, USA

[*]sborisk@mit.edu , gchen2@mit.edu

**Abstract**

We describe strategies to estimate the upper limits of the efficiency of photon energy harvesting via hot electron extraction from gapless absorbers. Gapless materials such as noble metals can be used for harvesting the whole solar spectrum, including the visible and the near-infrared light. The energy of photo-generated non-equilibrium or 'hot' charge carriers can in turn be harvested before they thermalize with the crystal lattice via the process of their internal photo-emission (IPE) through the rectifying Schottky junction with a semiconductor. However, the low efficiency and the high cost of noble metals necessitates the search for cheaper abundant alternative materials, and we show here that carbon can serve as a promising IPE material candidate. We compare the upper limits of performance of IPE photon energy-harvesting platforms, which incorporate either gold or carbon as the photoactive material where hot electrons are generated. Through a combination of density functional theory, joint electron density of states calculations, and Schottky diode efficiency modeling, we show that the material electron band structure imposes a strict upper limit on the achievable efficiency of the IPE devices. Our calculations reveal that graphite is a good material candidate for the IPE absorber for harvesting visible and near-infrared photons, whose electron density of states yields a sizeable population of hot electrons with the energies high enough to be collected across the potential barrier. We also discuss the mechanisms that prevent the IPE device efficiency from reaching the upper limits imposed by their material electron band structures. The proposed approach is general and allows for efficient pre-screening of materials for their potential use in IPE energy converters and photodetectors within application-specific spectral windows.

## 1. Introduction

Harvesting solar energy by photon absorption in metal nanostructures and subsequent collection of photo-generated hot electrons via the processes of internal photo-emission has been actively explored as an alternative approach to traditional photovoltaic (PV) as well as for catalysis and photo-detection [1]–[6]. The energy of the absorbed photons raises the energy of electrons above the Fermi level in the absorber material, creating a population of energetic, or 'hot' electrons, whose energies are out of thermal equilibrium with the crystal lattice. Hot electrons typically cool down very fast due to scattering on cold electrons, lattice defects, and phonons. The cooling process occurs on picosecond timescale in most metals [2], [4], [7]– [9]. If, however, these hot electrons can be extracted before they cool down, they can generate voltage and/or current in the external circuit [10].

One possible hot electron extraction scheme is based on their injection above the Schottky barrier that forms at the interface between a metal and a semiconductor (Fig. 1). In this scheme, hot electrons generated by absorption of photons with energies below the semiconductor bandgap can still be harvested [11]– [13]. This offers the way to potentially increase the conversion efficiency of PV cells and to extend the bandwidth of photo-detectors. Unlike conventional PV cells, which rely on the minority-carrier transport in semiconductor material, in Schottky devices the semiconductor is only used for the majority carrier transport and separation. The photo-generated hot electrons can also be used to drive catalytic reactions [3], [4], [14].

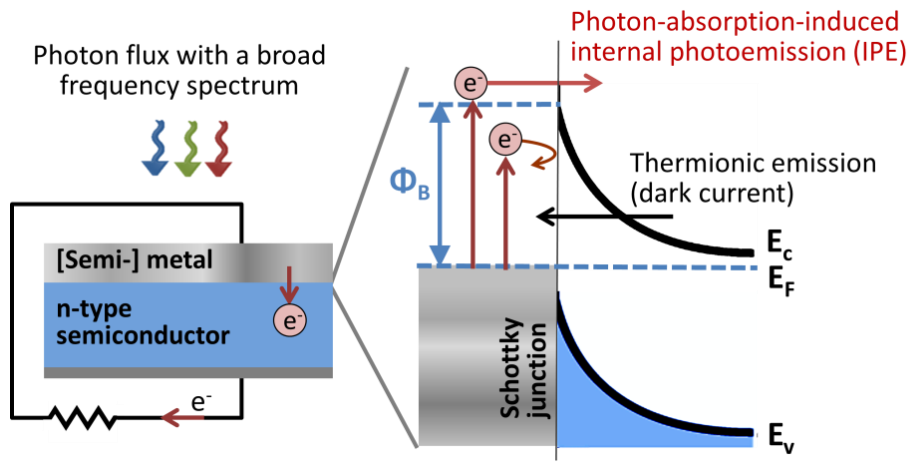

**Figure 1.** Schematic of the photon-to-electrical current energy conversion platform combining either a metal or a semimetal photon absorber and the Schottky barrier as the hot-carrier filter. Hot electron internal photoemission process across the Schottky junction between a (semi-)metal and an *n*-type semiconductor is also schematically shown.

Traditionally, plasmonic metals such as gold (Au) are considered as good material candidates for gapless photon absorbers, which are potentially capable of the full solar spectrum harvesting. Unfortunately, the photon-to-electricity energy conversion efficiencies in the hot-electron solar harvesting devices demonstrated to date have been extremely low [15]– [18]. The disappointing efficiency partially stems from the difficulty in achieving efficient and broadband light absorption in metal nanostructures. State-of-the art approaches to improving photon energy harvesting typically focus on the photon density of states engineering to achieve perfect light absorption in various materials. Efficiency and bandwidth of noble-metal absorptance can be increased, although it often requires precise nano-patterning and/or external optical trapping schemes. However, it has been recently revealed that the number of energy states available for electron transitions in the absorber material imposes strict limitations on the upper limits of efficiency achievable in IPE solar energy conversion platforms even if the perfect absorptance is achieved [19]– [22]. Accordingly, to select a good material candidate for a target spectral window, material properties beyond the ability to efficiently absorb photons should be considered. Another obvious limitation of gold and other noble metals is their high price stemming from low abundance of these materials on our planet. To overcome this limitation, other cheaper metals such as e.g., aluminum, copper, and silver, have been explored for applications in energy harvesting through hot electron generation [8], [22], [23]. Semimetals, such as e.g. carbon-based materials, lack an electronic bandgap, and can be used instead of metals for harvesting both visible and near-infrared photons. They can also form rectifying Schottky junctions with conventional semiconductor materials [24]– [28], making them promising material candidates for the IPE absorbers. In contrast to metals, which have a partially filled conduction band, semimetals are characterized by an overlap between the bottom of the conduction band and the top of the valence band. Unlike other semimetals such as arsenic, antimony, bismuth, and α-tin, some forms of carbon are non-toxic, cheap, and abundantly present in the earth core. Here, we consider two forms of carbon: graphite, which is a typical semimetal, and a monolayer graphene, which is a semimetal with a negligible density of states at the Fermi level, and in which the energy is proportional to the momentum rather than its square. In the following, we use *ab initio* calculations of the electron density of states in the absorber material to evaluate the upper efficiency limits of IPE energy harvesting platforms. We use gold and carbon as sample material candidates to evaluate and compare by using this approach, and demonstrate higher efficiency limits for the solar spectrum harvesting of sunlight potentially achievable with the use of graphite.

**2. The theory and modeling of the internal photoemission process**

A schematic of a typical IPE device is shown in Fig. 1, and consists of a metal or semi-metal photon absorber, which forms a rectifying Schottky junction at an interface with a semiconductor [29]– [31]. An electronic band diagram for the Schottky junction is shown in the right panel of Fig. 1 for the case of the junction between a metal and an n-type semiconductor. In the following, we limit the analysis to the hot-electron-based IPE devices. However, the theory is fully applicable to the hot-hole-type IPE photon energy harvesters that utilize Schottky junctions between (semi-)metals and p-type semiconductors [21], [32]. The electronic band gap in the semiconductor material forces formation of the energy barrier of height $\Phi_B$ at the material interface, which only allows the transport of hot electrons from metal to semiconductor with energies at least $\Phi_B$ above the Fermi level $\mathrm{E}_F$, as shown in Fig. 1. The height of the Schottky barrier depends on a combination of the metal and semiconductor materials, and typically varies in 0.5– 1.5 eV range [24], [26], [31], [33]. Once these hot electrons are injected into the semiconductor, the band gap prevents their recombination with holes and preserves their extra energy in excess of the Fermi level. While the non-equilibrium IPE process drives the forward current through the Schottky junction, the equilibrium thermionic emission from the semiconductor into the (semi-)metal results in the reverse dark current, which reduces the efficiency of the IPE photon-energy-harvesting platforms.

The principle of operation of the IPE device is based on the three-step process, involving (1) hot carriers generation by a combination of direct photon absorption and an indirect absorption via plasmon excitation and subsequent decay, (2) hot carriers transport to the interface, and (3) their injection into the semiconductor material [19]–[21], [34]. The two latter processes must compete with the hot carriers thermalization between themselves and with the lattice, which may reduce the IPE yield. For many decades, the semi-classical Fowler theory provided the analytical tool to evaluate the IPE yield. This theory is based on the spherical Fermi surface approximation for electrons in metals and the assumption that the kinetic energy of electrons *normal* to the barrier must exceed the barrier height. The above assumptions result in the prediction for the small angular escape cone for electrons, whose angular width is limited by the cut-off value of the hot electron momentum normal to the interface [35]. However, recent studies have shown that the IPE device performance can significantly exceed the limits predicted by the Fowler equation [7], [20], [34], [36]. The observed differences were mostly attributed to the deviation from the escape-cone limitation, with the extra momentum provided via electron-phonon scattering, plasmon-to-hot-electron decay, and surface roughness of the material interfaces [20], [34]. These observations are in line with prior work on super-lattice thermoelectric devices, where the increased number of hot electrons participating in the conduction process was attributed to the non-conservation of lateral momentum during the interface emission process [37], [38]. The hot electron yield increase becomes especially pronounced in the case of absorbers fully embedded within a semiconductor and having at least one dimension smaller that the electron mean free path limited by the electron-electron scattering [20], [34], [39]. Electron-phonon scattering increases the yield by changing the momentum of the non-equilibrium electrons and re-directing them into the escape cone without significant energy loss [40]. The above factors increase the probability of the hot electron reaching the interface and escaping into the semiconductor.

Some recent studies of the IPE devices employed simple band model – parabolic band approximation – for the electronic band structure of noble metal absorbers [19], [20]. However, such approximation fails to correctly predict the IPE efficiency limits for materials with non-parabolic bands [21], [23], [32], nano-scale absorbers exhibiting electron level quantization effects [7], and absorbers under a broadband light illumination, which causes both inter-band and intra-band photon absorption [21]. As such, the electronic band structure of the absorber material emerges as an important parameter that can either suppress or enhance the IPE yield in the Schottky-junction-type devices. The density of filled and empty electrons states in the absorber material, which are available for the photon-induced upward electron transitions, strongly affects the first step of the IPE process, and thus establishes the absolute upper limit to the IPE device efficiency.

In the following, we use the density functional theory (DFT) calculations of the available density of electron states in the absorber material and illustrate its effect on the upper efficiency limit of IPE devices. We use the DFT to calculate the electron band structure in gold and carbon materials, and predict significant differences in their upper efficiency limit estimates imposed by the materials' electronic band structures. The electronic band structures of graphene and graphite offer potential performance improvement of IPE devices for the full or partial solar spectrum harvesting.

**3. Band-structure-imposed upper limits of the IPE efficiency: gold versus carbon**

We calculate the upper limit of the IPE light-to-current conversion efficiency under the assumptions of (i) perfect photon absorption across the whole solar spectrum, (ii) ballistic hot electron transport to the Schottky junction between the (semi-)metal and the semiconductor, and (iii) momentum matching of the hot electrons that makes possible their transport through the potential barrier (i.e., no escape cone limitation) [19]– [21]. The short circuit current generated by light absorption in a Schottky device with the potential barrier of height $\Phi_B$ can be calculated as follows:

$$J_{sc} = q \cdot \int \eta_{abs}(\hbar\omega) \cdot I(\hbar\omega) \cdot \eta_{IPE}(\hbar\omega, \Phi_B) d\omega. \tag{1}$$

Here, $q$ is the electron charge, $\hbar$ is the reduced Planck's constant, $\hbar\omega$ is the energy of photon, $I(\hbar\omega)$ is the incoming photon spectral flux, and $\eta_{abs}(\hbar\omega)$ is the spectral photon absorptance. The IPE efficiency $\eta_{IPE}(\hbar\omega, \Phi_B)$ is calculated as the probability of the photo-generated electron to be injected across the potential barrier (see below). For solar harvesting applications, $I(\hbar\omega) = C \cdot I_{AM1.5D}(\hbar\omega)$, where $I_{AM1.5D}(\hbar\omega)$ is the AM1.5D (ASTM G173-03) terrestrial solar spectrum [41], and $C$ is the solar concentration. Perfect absorptance condition yields $\eta_{abs}(\hbar\omega) = 1$ across the solar spectrum frequency range.

Under the ideal conditions of the ballistic electron transport and momentum non-conserved interface transport – i.e., considering the initial non-equilibrium distribution of hot electrons and assuming all electrons with energies higher than the barrier will transmit through the interface, the upper limiting value for $J_{sc}$ is solely determined by the IPE efficiency $\eta_{IPE}(\hbar\omega, \Phi_B)$. IPE efficiency characterizes the fraction of hot electrons that are generated by absorption of light with frequency $\omega$, which have enough energy to be injected from the (semi-)metal into the semiconductor above the Schottky barrier [19], [21]. IPE efficiency strongly depends on the electron band structure of the material, absorbed photon energy, and the potential barrier height $\Phi_B$. It can be calculated as a ratio of the population of the hot electrons with sufficient energy to be emitted over the Schottky barrier to the total photo-excited electron population:

$$\eta_{IPE}(\hbar\omega, \Phi_B) = \int_{\Phi_B}^{\hbar\omega} D(E, \hbar\omega) \, dE \Big/ \int_0^{\hbar\omega} D(E, \hbar\omega) \, dE. \tag{2}$$

The population of photo-excited electrons in a given material is determined by its joint density of states, which depends on the number of initial $(\rho(E - \hbar\omega))$ and final $(\rho(E))$ electronic states available for upward transitions driven by photon absorption:

$$D(E, \hbar\omega) = \rho(E - \hbar\omega) \cdot f_{FD}(E - \hbar\omega) \cdot \rho(E) \cdot (1 - f_{FD}(E)). \qquad (3)$$

Here, $f_{FD}(E - \hbar\omega) = (\exp\{(E - \hbar\omega - E_f)/k_b T\} + 1)^{-1}$ is the Fermi-Dirac distribution function, which defines the probability of the energy level $E - \hbar\omega$ to be occupied at a the initial equilibrium lattice temperature, while $(1 - f_{FD}(E))$ defines the probability of the energy level $E$ to be empty and available for the upward electron transition in the process of photo-excitation, $k_b$ is the Boltzmann constant, and $T$ is the lattice temperature.

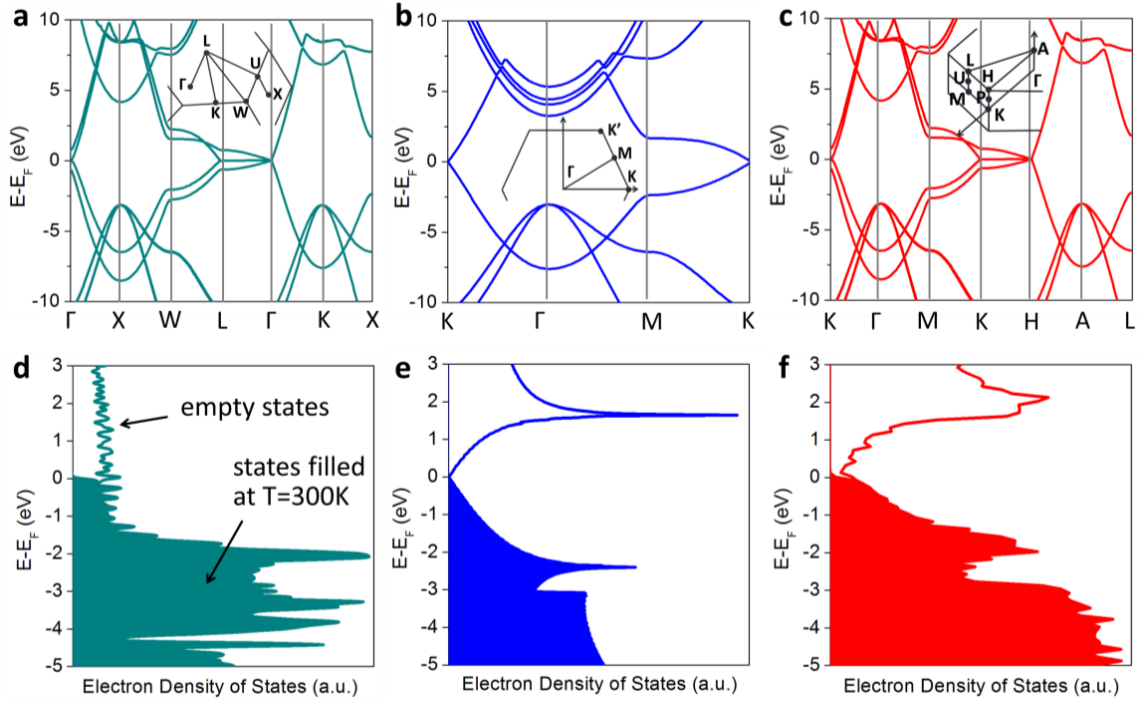

**Figure 2.** (a-c) Electron band structures of gold (a), graphene (b) and graphite (c) calculated by using the first-principles DFT method. The insets show the irreducible wedges of the Brillouin zones and the high-symmetry points in the reciprocal space of each crystal lattice. (d-f) The electron population of available energy states in the dark at T=300K (filled curves) and the density of electron states available for the upward electron transitions (empty curves) for Au (d), graphene (e) and graphite (f).

We calculate the electron band structure and the density of available electron states via the DFT simulations with the QUANTUM ESPRESSO package [42]. We use the local density approximation (LDA) of Perdew and Zunger [43] for all the materials examined, including gold, graphene and graphite. The projector augmented wave method [44] is used for gold while pseudopotential given by Hartwigsen-Goedecker-Hutter [45] is applied to graphite and graphene. Spin-orbit coupling in a scalar-relativistic level is also included for gold. Atomic coordinates in all cases are relaxed until the minimum force is below $5 \times 10^{-4}$ [Ry/au]. Electron band structures of the three materials are shown in Figs. 2a-c. The electron band energies were subsequently interpolated onto a much finer mesh using the tetrahedra method [46] to calculate the electron density of states for the three materials shown in Figs. 2d-f. Detailed parameters used for each material are given in Table 1.

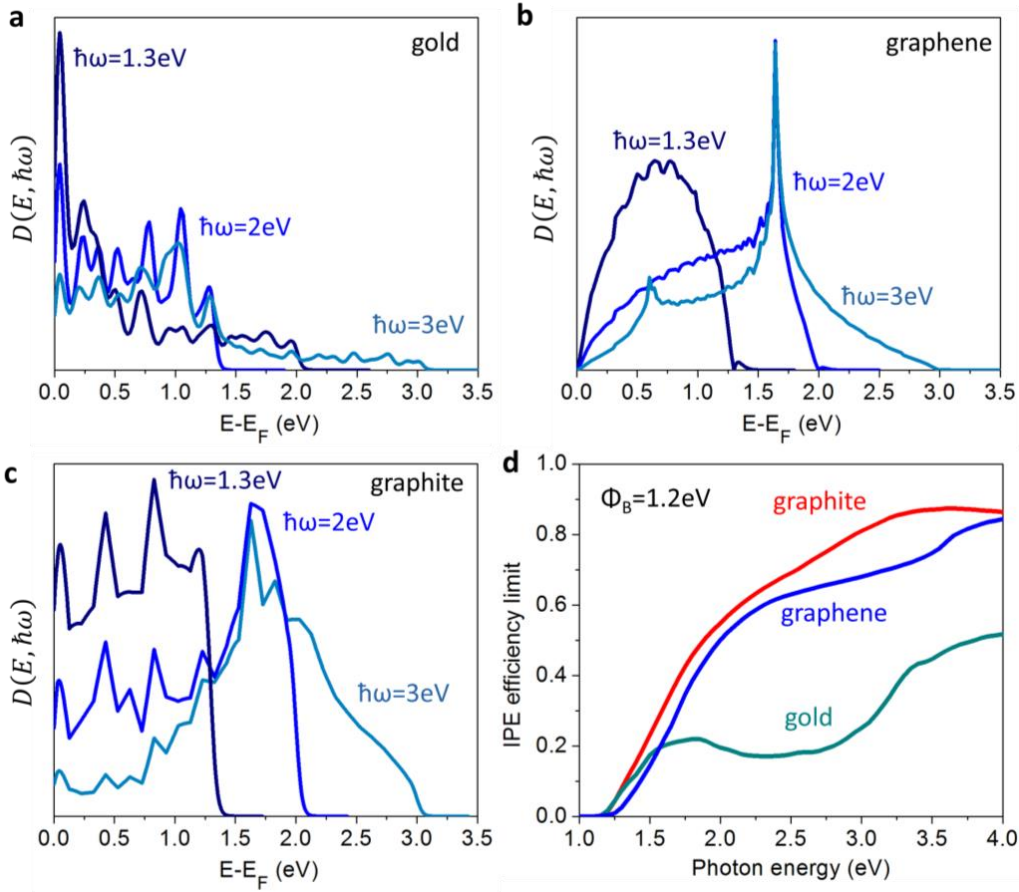

**Figure 3.** (a-c) Joint density of states $D(E, \hbar\omega)$ in Au (a), graphene (b) and graphite (c) for the photon energies $\hbar\omega = 2$eV (blue lines), $\hbar\omega = 1.3$eV (navy lines), and $\hbar\omega = 3$eV (light blue lines). (D) Limiting IPE efficiencies as a function of the absorbed photon energy for a fixed barrier height $\Phi_B = 1.2$eV (teal: Au, blue: graphene, red: graphite).

In Figs. 2d-f, we show the population of electron states filled at room temperature ($\rho(E - \hbar\omega) \cdot f_{FD}(E - \hbar\omega)$, filled curves) and the density of electron states available for the upward electron transitions in all the three materials considered. As can be seen in Fig. 2, the electron density of states calculated with the ab-initio DFT simulations deviates significantly from the parabolic band approximation used in some previous works [19], [20]. Instead, electron densities of states for different materials exhibit complex energy distributions that peak at different energies relative to the Fermi level in the material.

**Table 1.** Parameters used in the DFT calculations of the electron density of states.

| Material | Graphite | Graphene | Gold |
|---|---|---|---|
| Cut-off energy (Ry) | 60 | 80 | 80 |
| Gaussian broadening parameter (Ry) | 0.005 | 0.005 | 0.005 |
| **k**-mesh (before interpolation) | 20x20x12 | 20x20x1 | 25x25x25 |

Figures 3a-c compare the joint density of states ($D(E, \hbar\omega)$) calculated by using Eq. 3 for the photon energy at the peak of the solar radiation $\hbar\omega = 2$eV (blue lines) as well as for two other energies within the solar spectrum, $\hbar\omega = 1.3$eV (navy lines) and $\hbar\omega = 3$eV (light blue lines) for gold (a), graphene (b), and graphite (c). In Au, the population of hot electrons is dominated by the photo-induced transitions from the d-bands, which results in creation of many 'lukewarm' electrons with energies just above the Fermi level (Fig. 3a). Since only hot electrons with energies higher than the Schottky barrier height can be internally emitted into the semiconductor, the photo-excited hot electron population in Au yields low IPE efficiency. The joint densities of states of two carbon materials both exhibit peaks at higher hot electron energies (Figs. 3b,c), with a larger portion of hot electrons occupying higher-energy final states created in graphite than in graphene. Assuming that only hot electrons with energy higher than the barrier height can be internally emitted into the semiconductor over the potential barrier with a non-negligible probability, we use Eq. 2 to calculate the limiting IPE efficiency as a function of the photon energy and barrier height. This efficiency is plotted in Fig. 3d as a function of the energy of absorbed photons for all the three materials for a fixed barrier height $\Phi_B = 1.2$eV. The data in Fig. 3d predict the highest IPE efficiencies for graphite at all frequencies of the solar spectrum, leading to the conclusion that the electron density of states in graphite is most favorable for achieving high IPE efficiency among the three materials considered.

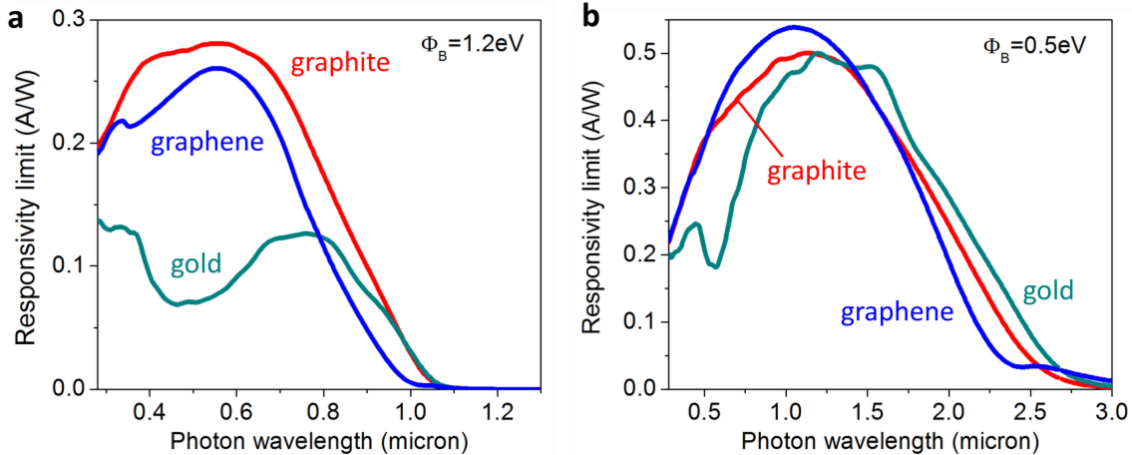

**Figure 4.** Responsivity limits for the IPE photo-detectors with gold (teal), graphene (blue), and graphite (red) as the absorber materials as a function of the photon wavelength for two potential barrier heights, $\Phi_B = 1.2$eV (a) and $\Phi_B = 0.5$eV (b).

By substituting the data in Fig. 3 to Eq. 1, we can now calculate the responsivity limit of the IPE photodetectors with different absorber materials as the ratios of their photo-generated short-circuit currents at zero applied voltage per watt of incident radiant power. This limit is plotted in Fig. 4 as a function of the incoming photon wavelength for two different Schottky barrier heights of 1.2eV and 0.5eV, respectively. Once again, we can observe the strong effect from the absorber material electron density of states on the IPE device responsivity limit, with carbon materials out-performing gold across the whole visible spectrum for any potential barrier height. The Au device responsivity limit approaches those for carbon-based IPE devices in the near-IR range, and may even exceed them in the case of low-height potential barrier (Fig. 4b).

However, the photo-generated current that can be delivered to the external load is reduced by the reverse dark current due to thermionic emission through the Schottky barrier from the semiconductor into the (semi-)metal. Total current is defined trough the standard Schottky diode equation as follows:

$$J = J_{sc} - A_R \cdot T^2 \cdot e^{-\Phi_B/k_bT} \cdot \left(e^{V/k_bT} - 1\right), \tag{4}$$

where $V$ is the applied voltage and $k_b$ is the Boltzmann constant. The thermionic emission reverse current scales inverse exponentially with the potential barrier height, as the second power with the temperature, and depends on the specifics of the materials interface through the modified Richardson constant $A_R$. In the following calculations, we use the Richardson constant value for titanium dioxide ($A_R = 6.71 \times 10^6 Am^{-2}K^{-2}$), which was previously reported in the literature [47]. Titanium dioxide ($TiO_2$) is a common choice for semiconductor used in IPE devices [6].

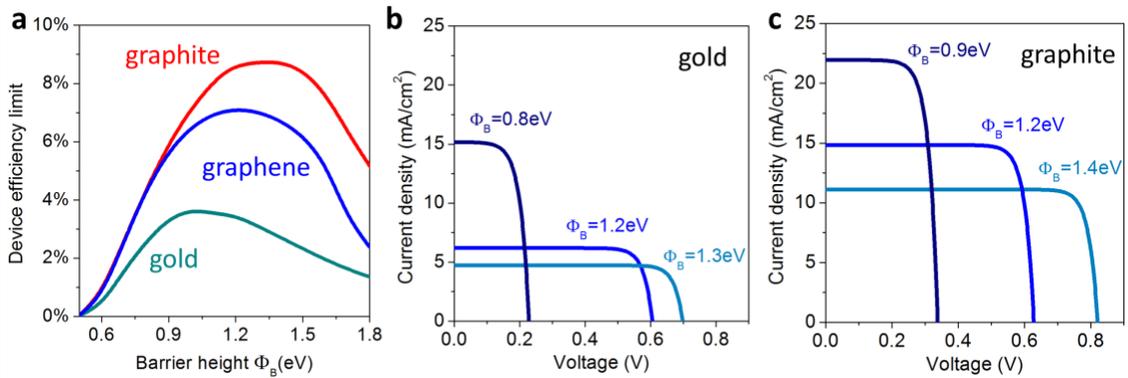

**Figure 5.** (a) Limiting IPE efficiencies as a function of the potential barrier height (teal: Au, blue: graphene, red: graphite). (b,c) Typical I-V curves for different barrier heights (shown as labels) for gold (b) and graphite (c).

We define the overall limiting efficiency of the IPE solar energy converter as the ratio of the maximum electrical power delivered to the load to the total power of the incoming sunlight, i.e.:

$$\eta = max(J \cdot V)/I_{in}, \tag{5}$$

$$I_{in} = C \cdot \int_{solar} I_{AM1.5D}(\hbar\omega)d\omega. \tag{6}$$

The maximum power point $max(J \cdot V)$ can be found by solving the following equation:

$$d(J \cdot V)/dV = 0. \tag{7}$$

The upper limits of the overall conversion efficiency calculated by using Eqs. 4-7 for Au, graphene and graphite are plotted in Fig. 5a as a function of the Schottky barrier height. All the plots have a characteristic bell shape, exhibiting efficiency reduction for low barrier heights due to the enhanced thermionic reverse current, and a similar drop at high barrier heights due to the reduced number of hot electrons with high enough energy to cross the potential barrier. Typical I-V curves for several barrier heights are shown in Figs. 5b,c for gold and graphite devices, and illustrate the common trend of the short circuit current increase (decrease) at the expense of reduced (increased) open circuit voltage at low (high) barrier heights. The I-V curves for graphene look very similar and were omitted for brevity. The data in Fig. 5a predict that graphite can potentially offer over twofold increase in the overall device efficiency over Au, reflecting the larger population of energetic electrons photo-generated in graphite (Fig. 2) and its corresponding higher IPE efficiency (Fig. 3). The optimum barrier height for the graphite-based device is higher than those for the Au- and graphene-based ones, but graphite is expected to out-perform the other materials at any barrier height. The efficiency limit increase of the graphite-based IPE device over that of graphene-based one is not as significant as over the Au-based platform. However, this limit assumes 100% light absorption across the broad solar spectrum, and thus the actual efficiency is expected to be much higher for graphite than graphene IPE devices due to significant difficulties in achieving efficient broadband absorption in 2D graphene.

**4. Increasing efficiency levels with solar concentration and spectral splitting strategies**

Similar to the Shockley-Queisser limit of the photovoltaic cells, the conversion efficiency of the IPE converter with the graphite absorber can be further increased by using either solar concentration or spectral splitting (or both) [1], [48]. Sorting solar photons by their energies and processing different parts of the solar spectrum separately by using several conversion platforms is a well-known way to increase the light-to-current energy conversion efficiency [1], [49], [50]. Single-junction PV cell can only convert efficiently photons with energies within the so-called PV band just above its electron bandgap [1], [51], [52]. Higher-energy photons are absorbed efficiently in the cell yet their energy is partially lost during the charge-carrier thermalization process [53], [54]. In turn, photons with low energies are not harvested by the PV cell at all. To achieve high overall energy conversion efficiency, PV cells can be incorporated into hybrid spectral-splitting energy conversion platforms [1], [49], [51], [52], where high- and low-energy photons are processed by different converters.

In Fig. 6 we evaluate the possibility of using the graphite IPE device as a part of the hybrid solar energy conversion platform, which also incorporates a conventional photovoltaic cell. Figure 6a schematically shows one of many possible ways of splitting the solar spectrum into three parts: high-energy band, low-energy band, and the central PV band. The data on Fig. 6b show that conversion of only high-energy and low-energy parts of the solar spectrum can be performed at efficiency higher than that of the full-spectrum conversion (compare solid blue and light blue lines in Fig. 6b). The PV band of the solar spectrum can then be directed to a conventional photovoltaic cell with the help of selective mirrors or filters [1], [51] to be converted at high efficiency. It should be noted that in this case the PV cell efficiency will also exceed its Shockley-Queisser [54] limit owing to elimination of the low- and high-energy parts from the photon spectrum [1], [55]. The high-energy part of the solar spectrum alone (solid purple line in Fig. 6b) can be converted by the graphite IPE converter at much higher efficiency than the whole broadband solar spectrum (solid blue line in Fig. 6b). Even the low-energy part of the solar spectrum can be converted to electricity with ~4% efficiency by a graphite-based IPE device (solid navy line in Fig. 6b). These results indicate a promising way of using the IPE converters as a part of hybrid solar harvesting and conversion platforms, especially at higher optical concentrations.

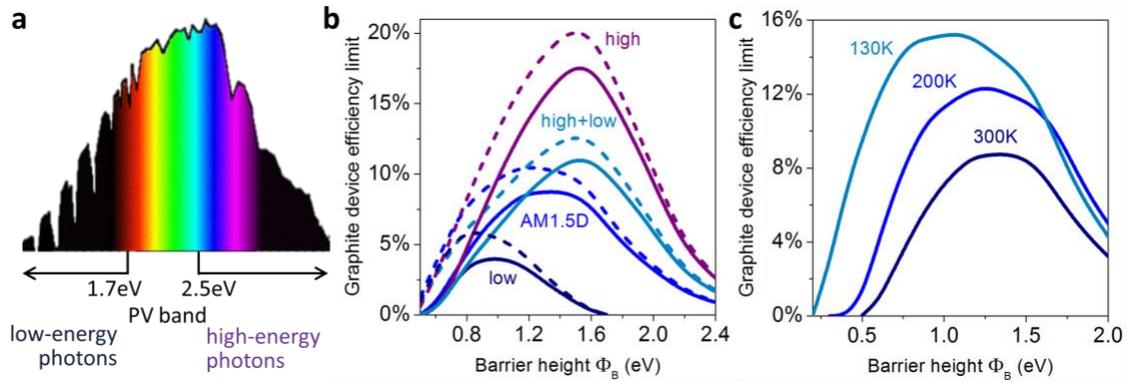

**Figure 6.** (a) The spectral power distribution in the AM1.5D solar spectrum, and one of many possible ways to achieve its spectral splitting into high-energy, low-energy and PV spectral bands. (b) Limiting energy conversion efficiencies of the graphite IPE converter as a function of the potential barrier height under partial solar spectrum illumination and varying solar concentration, including: the efficiency of the full solar spectrum conversion (blue) as well as efficiencies of the partial spectrum conversion of high-energy photons ($\hbar\omega > 2.5 eV$, purple lines), low-energy infrared photons ($\hbar\omega < 1.7 eV$, navy lines), and both, high and low-energy parts of the spectrum (light blue lines). Solid curves are for the solar concentration of one sun, dashed lines are for the solar concentration of 100 suns. (c) Limiting energy conversion efficiency of the graphite IPE converter for different temperatures of the device (shown as labels) and solar concentration of 100 suns.

The conversion efficiency of the IPE converter with the graphite absorber can be further increased by concentrating sunlight with external optics to increase photon flux [1], [48]. Our calculations predict that for the illumination with concentrated sunlight of 100 suns, the efficiency maximum for the full spectrum processing increases and exceeds 10% (dashed blue lines in Fig. 6b). The efficiency plot retains its bell shape, with the maximum point shifting towards lower barrier height. Solar concentration increases device efficiency limit for any spectral band, as can be seen by comparing the dashed lines (100 suns concentration) to solid lines (1 sun) of the same color for different spectra windows in Fig. 6b.

Temperature of the device is another parameter that can be tuned to improve its overall energy conversion efficiency. Since reverse thermionic current scales as the second power with temperature (Eq. 4), low-temperature operation would increase the performance of the IPE device. This is illustrated in Fig. 6c, where the graphite IPE device efficiency for the full solar spectrum conversion is plotted as the function of the potential barrier height at different temperatures. It can be seen that the maximum efficiency of the graphite IPE device at 130K is almost double of that at 300K, making it a potentially promising platform for airborne and space applications.

## 4. Discussion and conclusions

The above upper-limit efficiency analysis was performed under an assumption that all the solar energy can be absorbed by the (semi-)metal material in the IPE device, and that the photo-excited hot electrons can reach the barrier before they cool down [39], [56]. In reality, a fraction of the photo-excited electrons will cool down before reaching the barrier. Hot carrier cooling occurs though a combination of electron-electron and electron-phonon scattering processes [7], [57]. However, if the absorber thickness is comparable to or smaller than the hot carrier mean free path in the absorber material, multiple reflections of hot electrons from the surfaces of the thin film can take place, increasing the probability of the IPE process [39]. It is known from prior measurements and DFT modeling that mean free path for hot electrons in noble metals such as Au range from a few nanometers for electrons with energies exceeding the Fermi level by 2-5 eV to 50-100 nm for electrons with energies close to the Fermi level [9]. Recent DFT modeling predicts comparable mean free path values for higher-energy electrons in graphene and graphite, and significantly larger ones (from ~100nm to above 1000nm) for electrons with the energies within 0.5eV above the Fermi level [58]. Furthermore, unlike in noble metals, electron-phonon scattering is predicted to be the dominant scattering process for the non-equilibrium electrons, especially for those with higher energies. This effect was attributed to softer phonon modes and stronger electron–phonon coupling with lighter atoms in carbon materials than in metals [58]. The above observations hold high promise for the use of hot-carrier carbon IPE devices as photodetectors for both visible and infrared spectral ranges [12], [59].

Graphene has already been actively explored for hot-electron harvesting applications [60]– [62]. However, achieving efficient broadband absorption in graphene is significantly more challenging than either in gold or in graphite [63]. The use of graphite offers carbon-based efficient and broadband light harvesting as well as electron characteristics more favorable for generating high-energy photo-excited charge carriers, which translates into higher IPE device efficiency. Schottky-junction IPE energy converters with graphite as the active absorber material can be fabricated with a variety of semiconductors, including Si, $TiO_2$, SiC, GaAs, etc., allowing for tuning and optimization of the potential barrier height [24], [28]. Finally yet importantly, the price of graphite is orders-of-magnitude below that of gold owing to it much higher abundance in the earth core. While graphite costs about 1.5K US dollars per metric ton, gold costs about 40K US dollars per kilogram [64].

Overall, our data highlight the strong effect the material's electron density of states structure can have on the efficiency of the solar-harvesting IPE devices. Our results also illustrate that the electron density of states in graphite together with the long hot electron mean free path, low price, and abundance makes it a promising IPE absorber material candidate for harvesting visible and near-infrared photons. The technique described in this paper offers a useful strategy to screen other potential material candidates for use as absorbers in IPE photo-detectors and photon energy harvesters. It can also be used to engineer new nano-structured materials (including carbon materials) with tailored electron density of states that maximizes the device efficiency. An ideal absorber material that would maximize the high-energy hot electron population should exhibit a high and narrowband peak in its DOS just below the material Fermi level [19]. The energy distribution of the graphite DOS is much close to this ideal situation than that of Au, which contributes to the efficiency limit improvement predicted in this work. Further material improvements can be accomplished by using quantum confinement effects in heterostructures and alloys.

**Acknowledgements**

This work has been supported by the US Department of Energy, Office of Science, and Office of Basic Energy via 'Solid State Solar-Thermal Energy Conversion Center (S3TEC)', Award No. DE-SC0001299/DE-FG02-09ER46577. We thank Gerald Mahan, Wei-Chun Hsu, Mildred Dresselhaus, Prineha Narang, and Ravishankar Sundararaman for useful discussions and data sharing.

**Authors' contributions:** S.V.B. conceived the idea, J.Z. and Z.D performed DFT modeling, S.V.B. performed IPE efficiency calculations, G.C. and S.V.B. supervised the project, all the authors contributed to discussions and manuscript writing.

The authors declare no conflict of interest.